\documentclass[fp,twocolumn]{jpsj3}

\usepackage[dvipdfmx]{graphicx}
\graphicspath{{../}} 
\usepackage{amsmath}
\usepackage{amsfonts}
\usepackage{dcolumn}
\usepackage{txfonts}
\usepackage{bm}
\usepackage{physics}
\usepackage{mathtools}
\usepackage{ascmac}
\usepackage{color}
\usepackage{graphics}
\usepackage{float}

\title{Theoretical Study of Impurity Effects on Superconductivity in UTe$_2$}

\author{Koki Doi$^1$, Shingo Haruna$^1$, Mutsuki Iwamoto$^1$,Takuji Nomura$^{1,2}$, and Hirono Kaneyasu$^{1,3}$}
\inst{$^1$Department of Material Science, University of Hyogo, 3-2-1 Kouto, Ako, Hyogo 678-1297, Japan \\
	$^2$Synchrotron Radiation Research Center, National Institute for Quantum Science and Technology, 1-1-1 Kouto, Sayo, Hyogo 679-5148, Japan \\
	$^3$Center for Liberal Arts and Sciences, Sanyo-Onoda City University, 1-1-1 Daigakudori, Sanyo-Onoda, Yamaguchi 756-0884, Japan}

\abst{
	This study investigates the impurity effects on $\mathrm{UTe_2}$ within the self-consistent Born approximation using the six-orbital $ f $-$ d $-$ p $ model which contains two uranium and tellurium atoms in the minimum unit cell. 
	We analyze the dependence of superconducting transition temperature ($T_{c}$) on impurity concentration for various pairing symmetries proposed by experiments and theories.
	It clarifies that the decrease of $T_{c}$ signiﬁcantly depends on which atom sites the impurities reside. 
  Particularly, the analysis shows that the impurity at U-site has a dominant effect on the change of $T_{c}$.
	Then, either the singlet state in the case of magnetic impurities or the triplet states in both nonmagnetic and magnetic impurities are consistent with experiments.
	Thus, this indicates that elucidating the magnetic properties of impurities (i.e. magnetic or nonmagnetic) is crucial for identifying the pairing symmetry of $\mathrm{UTe_2}$.
}

\begin{document}
\maketitle

\section{introduction}
 UTe$_{2}$ was first synthesized by the chemical vapor transport method as a uranium (U)-based heavy-fermion superconductor.\cite{Ran1}
Subsequently, high-quality samples synthesized by the molten salt flux (MSF) method reports the superconducting transition temperature $T_{c}$ of 2.1K.\cite{Sakai2, MSFLT}
The multiple superconducting phases are yielded by applying pressure and field, implying nontrivial spin and orbital degrees of freedom of the Cooper pairs.\cite{Lewin2023, Aokireview, Rosuel1, Knebel1, Sakai3, Aoki4, Thomas1, Aoki3, Ran2}
UTe$_{2}$ shows the characteristic field dependence of the superconducting state, especially the upper critical ﬁeld $H_{c2}$ exceeds the Pauli limit along all the crystallographic axes.\cite{Rosuel1, Knebel1, Sakai3, Aoki4}
The field-induced reentrance behavior is considered the evidence of spin-triplet superconductivity, and it is similar to those of URhGe and UCoGe.\cite{Ucom}
Nuclear magnetic resonance (NMR) measurements shows that the NMR Knight shift decreases in all crystalline directions, which indicates the spin-triplet $A_{u}$ state with a full gap under the assumption of $p$-wave pairing.\cite{Matsumura1,Kitagawa1}
Moreover, heat transport also suggests the full-gap state of $A_{u}$.\cite{Suetsugu}
Specific heat measurements indicate the existence of point node,\cite{Ran1} and ultrasound measurement indicates the $B_{2u}$ state based on comparing the presence and anisotropy of discontinuities in the elastic moduli with the distribution of the density of states on the Fermi surface.\cite{Theuss1}
Experiments on the penetration depth demonstrate that the chiral state $B_{3u} + iA_{u}$ is required to explain the position of the point nodes.\cite{Ishihara1}
In these experimental situations, the irreducible representation of the superconducting state in UTe$_{2}$ remains unidentified and an open question.

Determining the pairing symmetry and gap structure is an important issue in understanding the topological properties and the mechanism of the superconductivity in UTe$_{2}$. \cite{Henrik1,Ohashi,Tei}
Theoretical study based on spin ﬂuctuations evaluated by the random phase approximation shows that the spin-triplet states are stabilized in parameter regions, which yields anti-ferro and ferromagnetic spin-fluctuations in the periodic Anderson model.\cite{Hakuno1}
However, numerical analysis of the Eliashberg equation using the third-order perturbation theory (TOPT) demonstrates the stabilization of $A_{g}$ (an anisotropic $s$-wave state) in a spin-singlet pairing.\cite{Haruna1}
The gap produced by TOPT has a point-node-like structure and no sign change between two superconducting bands.
Thus, the superconducting symmetry of UTe$_{2}$ has not yet been identified in a manner consistent with both experimental and theoretical results.

Impurity effects have served as a crucial role in discussing superconducting pairing symmetries.\cite{ANDERSON1, Tsuneto, AG4, Haran1, Haran2, Balatsky2006, Korshunov2016, Onari1}
Recent experiments have shown that the reduction of $T_{c}$ in UTe$_{2}$ follows the Abrikosov-Gor’kov (AG) curve.\cite{Weiland1,MSFLT}
Additionally, experiment on thorium (Th) substitution, which is often used as a nonmagnetic substituent without $f$-orbitals in uranium-based compounds, shows that the reduction of $T_{c}$ can be fitted well with a linear function of Th concentration.\cite{Moir2025}
The main purpose of this study is to investigate possible pairing states agreeing with those experimental observations.
Specifically, using a multi orbital model of UTe$_{2}$, we investigate how magnetic and nonmagnetic impurities affect $T_{c}$ and gap structure for candidate pairing states of UTe$_{2}$; those are $A_{g}$ (anisotropic $s$), $B_{3u}+ iA_{u}$, $A_{u}$, and $B_{2u}$, under a zero magnetic field and ambient pressure.
The impurity effects are treated within the self-consistent Born approximation (SCBA).\cite{Haran1, MineevSamokhin1999}
Moreover, some studies have demonstrated that uranium or tellurium (Te) site vacancies significantly affect superconductivity. \cite{Cairns2020, Haga2022}
To consider the degrees of freedom when introducing impurities to the unit cell, we treat impurity potentials in the orbital representation and define the impurity correlation coefficient as the probability of the simultaneous existence of different impurity sites in the unit cell.
The dependence of $T_{c}$ on the impurity correlation coefficient is analyzed for different impurity sites.

\section{Model and Formulation}

The unit cell of UTe$_2$ (space group: $Immm$) contains two U sites and four Te sites. In our model, two U sites with $f$ and $d$ orbitals and two Te sites (Wyckoff position: $4h$) with $p$ orbitals are taken into account.
We use the six-orbital $f$-$d$-$p$ model constructed within tight-binding approximation
\begin{align}
  \hat{H}_{0}(\bm{k})=\sum_{l_{1}l_{2}}\sum_{s}H_{0l_{1}  s l_{2} s }(\bm{k})c^{\dag}_{\bm{k}l_{1} s}c_{\bm{k}l_{2} s},
\end{align}
where $c_{\bm{k}l_{n}s}$ ($c^{\dag}_{\bm{k}l_{n}s}$) represents the annihilation (creation) operator at the momentum $\bm{k}$, orbital $l_{n}$ ($l_{n}$= $f$,$f'$,$d$,$d'$,$p$,$p'$) and spin $s$.
Here we should consider spin $s$  as pseudo-spin rather than bare spin  since the six-orbital $f$-$d$-$p$ model is constructed to reproduce the band structure incorporating spin-orbit coupling.
But hereafter simply we refer to $s$ as spin.
$H_{0}$ is the hopping matrix introduced in Ref.\cite{Haruna1,Ishizuka1}.
Diagonalizing the tight-binding Hamiltonian determines the quasi-particle energy $\xi_{a}(\bm{k})$ and diagonalization matrix $U_{l_{n}a_{m}}(\bm{k})$.
Both quantities are independent of spin since the hopping matrix $H_{0}$ is diagonal in spin space.
This model accurately represents the electronic state of UTe$_{2}$, consistent with the antiferromagnetic fluctuations near $Q=(0,\pi,0)$ observed in neutron scattering experiments \cite{Butch1, Duan1} and the de Haas-van Alphen effect. \cite{Aoki1, Aoki2}

Hamiltonian for impurity scattering is given by
\begin{align}
  \hat{H}_{imp}=\sum_{i}\sum_{l}\sum_{s_{1} s_{2}}u_{l s_{1} s_{2}}(\bm{r}_{i})c^{\dag}_{il s_{1}}c_{il s_{2}},
\end{align}
where $c_{ils_{1}}$ ($c^{\dag}_{il s_{2}}$) is the annihilation (creation) operator at the $i$-orbital $l$ and spin $s_{1(2)}$.

We consider the sum of the sites with impurities.
The impurity potential $u$ is:
\begin{align}\label{eq:imp_pot}
  u_{l s_{1} s_{2}}(\bm{r})
  =  v_{l}(\bm{r}) \delta_{s_{1} s_{2}}
  +w_{l}(\bm{r}) (\bm{S} \cdot \bm{\sigma})_{s_{1} s_{2}},
\end{align}
where $\bm{S}$ is the spin of the impurity and $\bm{\sigma}$ is the electron spin matrix.
$v_{l}$ is the nonmagnetic potential, and  $w_{l}$ is the magnetic one.

Concerning the magnetic properties of impurities, spin-orbit coupling allows pseudo-spin flip scattering to occur through a change in orbital angular momentum.
In the case of UTe$_{2}$, such scattering channels are likely to occur since several $f$-orbital components have a density of states at the Fermi level.\cite{Haruna1}
However, our model cannot fully treat this effect since we consider only one $f$-orbital per U site.
Instead, we consider pseudo-spin flipping effectively in the form of Eq.\eqref{eq:imp_pot}.
In this treatment, uranium vacancies may not naively be considered as nonmagnetic impurities, but as magnetic impurities.
As another possibility, small inclusion of magnetic byproducts formed during the MSF method can also act as magnetic impurities.\cite{Sakai2}
Thus, we consider both nonmagnetic and magnetic parts of impurity potential in this study.

\begin{figure}[t]
  \centering
  \includegraphics[width=0.95\columnwidth]{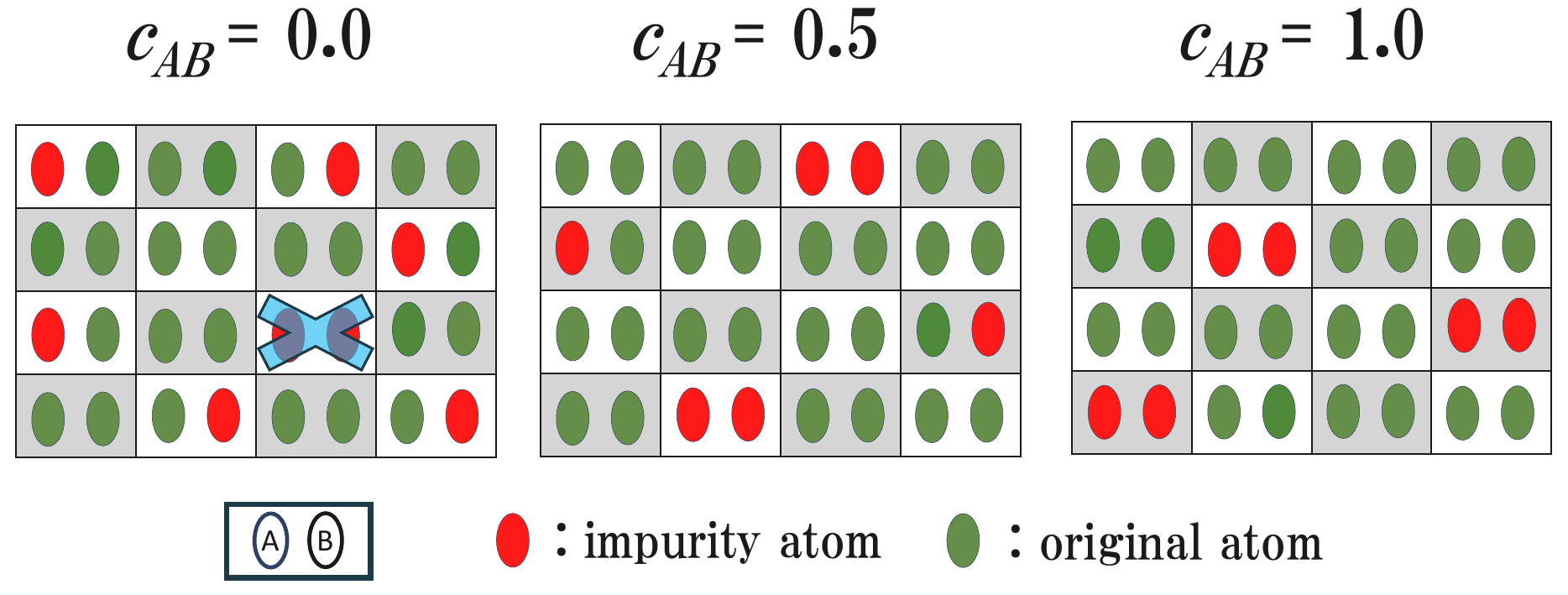}
  \caption{Typical cases of impurity configurations with $c_{AB}= 0.0, 0.5, 1.0$. When $c_{AB}=0.0$, the probability of having two impurity sites within a unit cell is neglected.}
  \vspace{-15pt}
  \label{fig1}
\end{figure}

By performing a random average over the spin of the magnetic impurities \cite{AG4} and assuming an on-site impurity potential, the effective scattering amplitude $I^{AB}$ in the orbital representation is obtained as follows:
\begin{align}
  I^{AB}_{l_{1} s_{1} l_{2}s_{2}, l_{2}s_{3} l_{1}s_{4}}=
   & c_{AB}\;n\Big[v^{A}_{l_{1}}v^{B}_{l_{2}}\delta_{s_{1} s_{4}}\delta_{s_{2} s_{3}}\notag                             \\
   & +\frac{1}{3}S(S+1)w^{A}_{l_{1}}w^{B}_{l_{2}}\bm{\sigma}^{A}_{s_{1} s_{4}} \cdot\bm{\sigma}^{B}_{s_{2} s_{3}}\Big],
\end{align}
where $S$ denotes the effective spin quantum number of the local impurities.         
We define the impurity concentration $n=$min$[n_{A},n_{B}]$, where $n_{A}$ and $n_{B}$
are concentration	at site A and B (U or Te).
For simplicity, we consider only the case $n_{A}=n_{B}$.
We introduce the impurity correlation coefficient $c_{AB}$, which quantifies the probability of impurity sites existing simultaneously at two different sites A and B in a unit cell, as shown in Fig.~\ref{fig1}.
For $c_{AB}$=1, if one of the A and B sites in a unit cell is an impurity site, then the other is always an impurity site.
For $c_{AB}$=0, A and B sites in a unit cell can be impurity sites unrelated to each other.
Here, for $c_{AB}$=0, we neglect the possibility that the A and B sites are both impurity sites in a unit cell, since such probability is of the order of $n_{A}n_{B}$.
$0<c_{AB}<1$ represents the fraction of impurity-containing unit cells in which $c_{AB}=1$.
Thus, the degree of scattering between impurity sites A and B is treatable.
Regarding scattering within the same impurity site, $c_{AA}=c_{BB}=1$.
For simplicity, we assume  the intersite magnetic correlations between impurity sites are absent, then in the magnetic-impurity cases we may restrict our calculation to the case of $c_{AB} =0$ $(A \neq B)$.

The normal self-energy within SCBA is given by
\begin{align} \label{eq:nse}
  \Sigma_{1a_{1}s_{1}a_{2}s_{2}}(k)
  =\sum\limits_{k^{'}}\sum\limits_{a_{3}a_{4}}\sum\limits_{s_{3}s_{4}}
  I^{N}_{a_{1}s_{1}a_{4}s_{4},a_{2}s_{2}a_{3}s_{3}}(\bm{k},\bm{k}^{'})
  G_{a_{3}s_{3},a_{4}s_{4}}(k^{'}),
\end{align}
where $k$ is a notation that combines the Matsubara frequency $\omega_{n}=(2n+1)\pi T$ and momentum $\bm{k}$.
$G$ is the normal Green’s function, determined by
\begin{align} \label{eq:ngf}
  G_{as,a^{'}s^{'}} & (k) = G^{(0)}_{as,a^{'}s^{'}}(k)\notag               \\
                    & ~~~+\sum\limits_{a_{1}a_{2}}\sum\limits_{s_{1}s_{2}}
  G^{(0)}_{as,a_{1}s_{1}}(k)\Sigma_{1a_{1}s_{1}a_{2}s_{2}}(k)G_{a_{2}s_{2},a^{'}s^{'}}(k).
\end{align}
Here, the bare normal Green's function $G^{(0)}$ is diagonal in both band and spin space.
Thus, we set $G_{as,a_{1}s_{1}}^{(0)}(k)=[i\omega_{n}-\xi_{a}(\bm{k})]^{-1}\delta_{a,a_{1}}\delta_{s,s_{1}}$.
By solving Eqs.~\eqref{eq:nse} and \eqref{eq:ngf} self-consistently, the normal Green's function and self-energy are determined.
Effective scattering rate $I^{N}$ in the band representation is given by
\begin{align} \label{eq:esr1}
  I^{N}_{a_{1}s_{1}a_{4}s_{4},a_{2}s_{2}a_{3}s_{3}}  (\bm{k},\bm{k}^{'})  =
  \sum\limits_{l_{1}l_{2}}&
  U_{l_{1}a_{1}}(\bm{k})U_{l_{2}a_{4}}(\bm{k}^{'}) U_{l_{2}\,a_{2}}(\bm{k})U_{l_{1}a_{3}}(\bm{k}^{'})                                                      \notag \\
                               & \times
  I^{AB}_{l_{1}s_{1} l_{2}s_{4}, l_{2}s_{2} l_{1}s_{3}},
\end{align}
where  $U_{l_{n}a_{m}}(\bm{k})$ represents a unitary matrix to diagonalize the six-orbital $f$-$d$-$p$ model.
We assume that the impurities are sufficiently dilute or have weak potential.
Thus, the introduction of impurities does not change the classification of bands and Fermi surfaces.
The normal self-energy is treated by neglecting its real part.
Although the effective scattering rate $I^{AB}$ in the orbital representation is momentum-independent, $I^{N}$ in the band representation acquires momentum dependence through the unitary matrix $U$.

\begin{figure}[t]
  \centering
  \includegraphics[width=0.95\columnwidth]{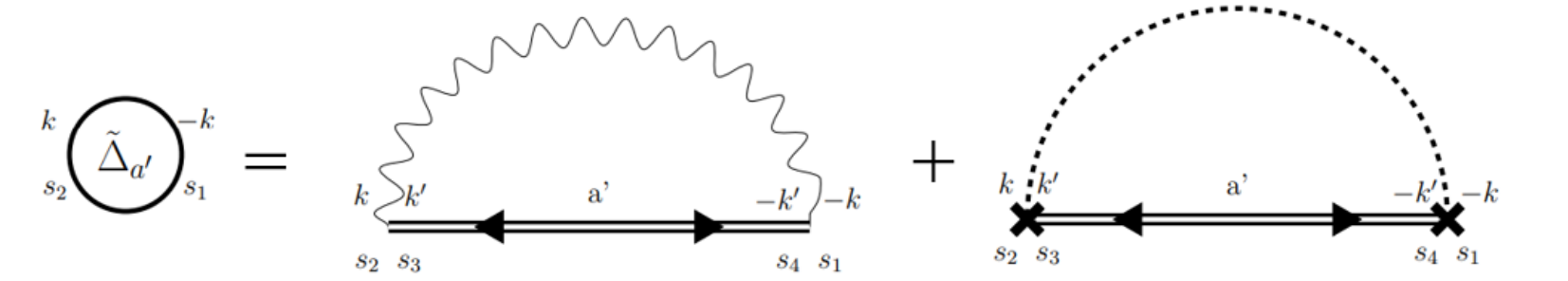}
  \caption{Feynman diagram of the BCS gap equation including the processes of impurity scattering. The first term on the right-hand side represents the original effective pairing interaction, while the second corresponds to the impurity effects within SCBA.}
  \vspace{-15pt}
  \label{fig2}
\end{figure}
To investigate the impurity effects of $T_{c}$ on UTe$_{2}$, we solve the linearized BCS gap equations as shown in Fig.~\ref{fig2}:
\begin{align}\label{BCS}
  \lambda\Delta_{as_{1}s_{2}}(k) & =
  T\sum\limits_{a^{'}}\sum\limits_{k^{'}}\sum\limits_{s_{3}s_{4}}
  \Bigg[g_{as_{1}s_{2},a^{'}s_{3}s_{4}}(k,k^{'})\notag                                              \\
                                 & ~~~+I^{S}_{as_{2}as_{1},a^{'}s_{4}a^{'}s_{3}}(\bm{k},\bm{k}^{'})
  \delta_{\omega_{n}\omega_{n}^{'}}/T\Bigg]  F_{a^{'}s_{3}s_{4}}(k^{'}),                            \\
  F_{a^{'}s_{3}s_{4}}(k)         & =
  \sum\limits_{a_{1}}\sum\limits_{s_{5}s_{6}}
  G_{a^{'}s_{3},a_{1}s_{5}}(k)\Delta_{a_{1}s_{5}s_{6}}(k)G_{a^{'}s_{4},a_{1}s_{6}}(-k),
\end{align}
where $\lambda$ is the eigenvalue, $\Delta$ is the renormalized superconducting gap, and $F$ is the anomalous Green's function.
In the square bracket of Eq.~\eqref{BCS}, the first term is the effective pairing interaction $g_{as_{1}s_{2},a^{'}s_{3}s_{4}}(k,k')=f_{as_{2}s_{1}}(\bm{k})f^{*}_{a's_{3}s_{4}}(\bm{k'})$, assumed to be separable without retardation effect.

\begin{fullfigure}[t]
  \centering
  \includegraphics[width=0.95\textwidth]{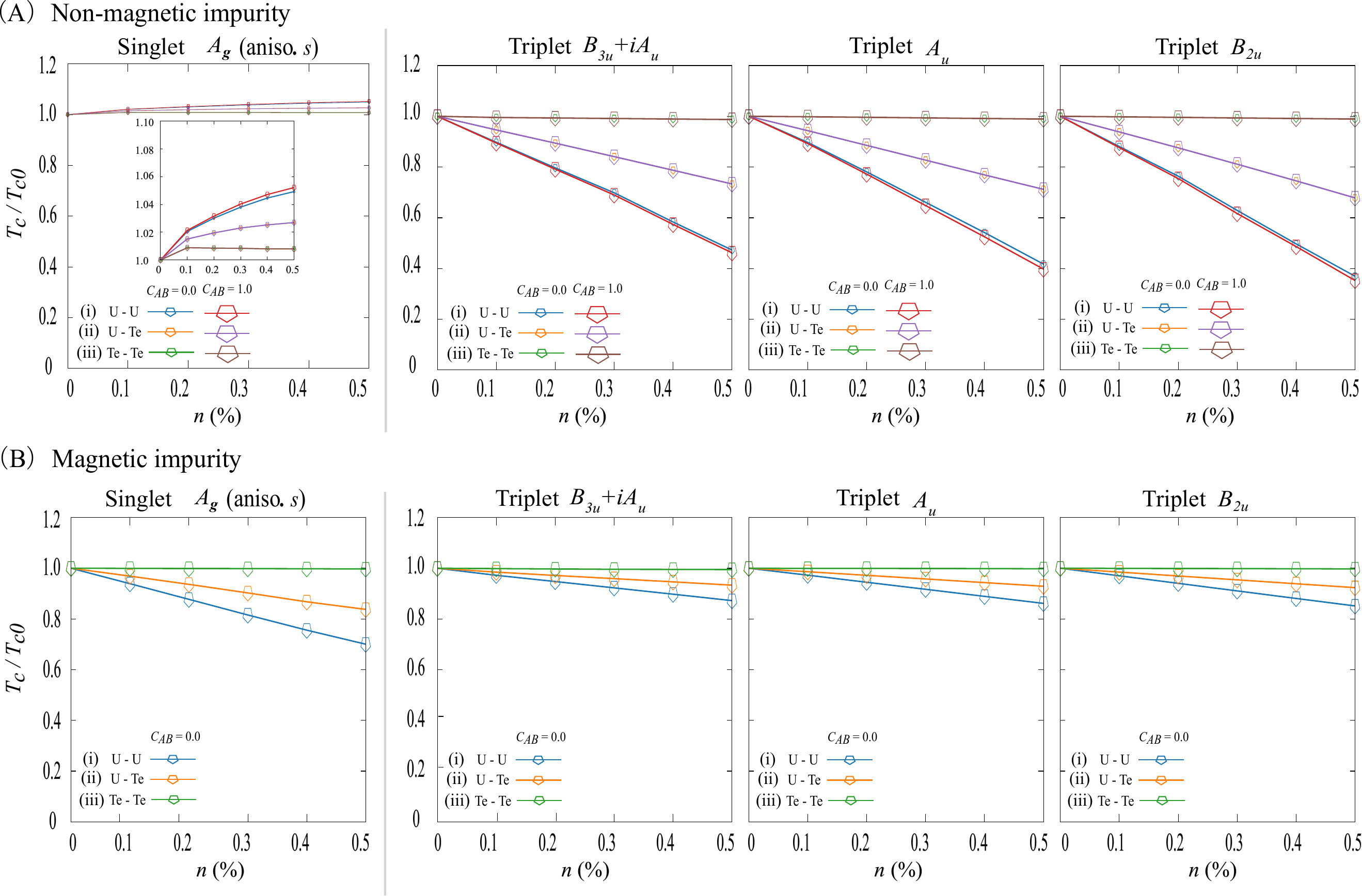}
  \caption{Change in $T_{c}$ as a function of the impurity concentration $n$ regarding (A) nonmagnetic  and (B) magnetic  impurities. For each superconducting symmetry, we considered three types of defects $[(\mathrm{i})$ U-U, $(\mathrm{ii})$ U-Te, and $(\mathrm{iii})$Te-Te$]$, with $c_{AB}=0,1$ (nonmagnetic impurities) and $c_{AB}=0$ (magnetic impurities).}
  \label{fig3}
\end{fullfigure}

The second term is the effective scattering rate $I^{S}$
\begin{align} \label{eq:esr2}
    I^{S}_{as_{2}as_{1},a^{'}s_{4}a^{'}s_{3}}(\bm{k},\bm{k}^{'})=
  \sum\limits_{l_{1}l_{2}}&
  U_{l_{2}a}(\bm{k})U_{l_{1}a}(-\bm{k})
  U_{l_{1}a^{'}}(-\bm{k}^{'})U_{l_{2}a^{'}}(\bm{k}^{'}) \notag \\
    & \times
  I^{AB}_{l_{2}s_{2}l_{1}s_{1}, l_{1}s_{4} l_{2}s_{3}}.
\end{align}
This term effectively plays a role of $s$-wave attraction and thereby for $s$-wave superconductors compensates the reduction of $T_{c}$ originating from the damping in the self-energy of Eq.~\eqref{eq:nse}, leading to the Anderson's theorem.\cite{ANDERSON1,Tsuneto}.
For non-$s$ wave pairings, ineffectiveness of the second term yields the significant reduction of $T_{c}$ following the AG theory.\cite{AG4}
$T_{c}$ is determined as the point at which $\lambda=1$ by solving Eqs.~\eqref{eq:nse}-\eqref{eq:esr2} self-consistently.

\begin{figure}[htbp]
  \centering
  \includegraphics[width=0.93\columnwidth]{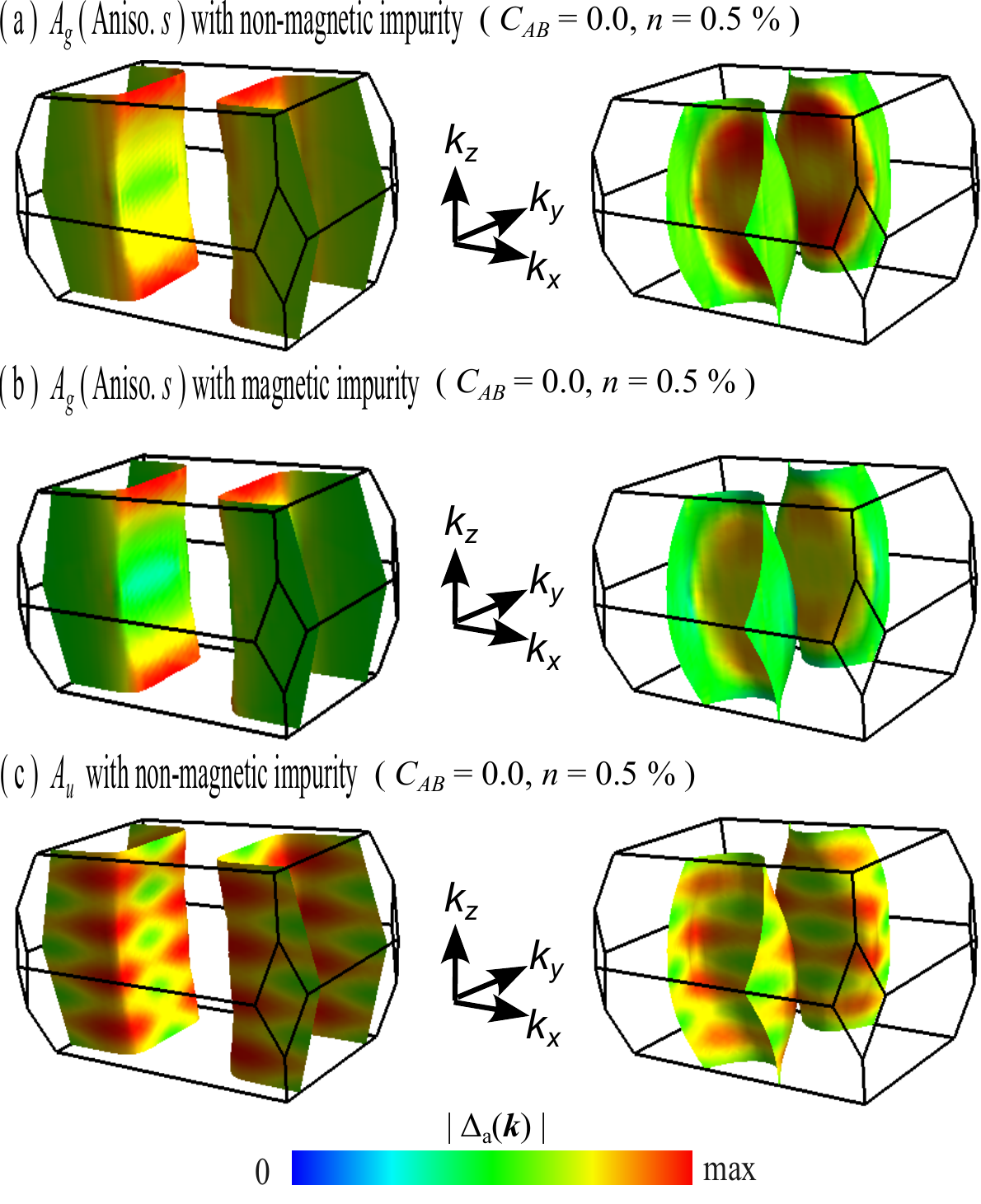}
  \caption{Superconducting gap structures $\lvert \Delta_{a}(\bm{k})\rvert$ for each pairing symmetry.
    (a) Gap structure for an anisotropic $s$-wave state with nonmagnetic impurities at $n=0.5\%$ and $c_{ab}=0$.
    (b) Gap structure for an anisotropic $s$-wave state with magnetic impurities at $n=0.5\%$ and $c_{ab}=0$.
    (c) Gap structure for $A_u$ state with nonmagnetic impurities at  $n=0.5\%$ and $c_{ab}=0$.}
  \label{Fig4}
\end{figure}

Analysis is performed using $24\times24\times24$ $\bm{k}$ meshes and 2400 Matsubara points.
For an anisotropic $s$-wave state, we use the $\bm{k}$-dependence of gap obtained by TOPT.\cite{Haruna1}
For the triplet states $(B_{3u}+ iA_{u}, A_{u}$, and $B_{2u})$, we impose the same gap amplitude on the two superconducting bands.
The $\bm{d}$ vectors are defined at each $\bm{k}$-point as follows:
\begin{align}
  \bm{d}_{A_{u}}(\bm{k})  & \propto \sin\big(k_{x}a\big)\hat{\bm{x}} + \sin\big(k_{y}b\big)\hat{\bm{y}} + \sin\big(k_{z}c\big)\hat{\bm{z}}, \notag \\
  \bm{d}_{B_{2u}}(\bm{k}) & \propto \sin\big(k_{z}c\big)\hat{\bm{x}} + \sin\big(k_{x}a\big)\hat{\bm{z}},                      \notag       \\
  \bm{d}_{B_{3u}}(\bm{k}) & \propto \sin\big(k_{z}c\big)\hat{\bm{y}} + \sin\big(k_{y}b\big)\hat{\bm{z}}, \notag
\end{align}
where $a$, $b$, and $c$ represent the lattice constants, and $\hat{\bm{x}}$, $\hat{\bm{y}}$, and $\hat{\bm{z}}$ represent the unit vectors in the spin space determined by the crystal axes.
We define $T_{c0}$ as the superconducting transition temperature in the absence of impurities.
The values are $T^{A_{g}}_{c0}=9.789$ meV, $T^{B_{3u}+iA_{u}}_{c0}=9.735$ meV, $T^{A_{u}}_{c0}= 8.633$ meV, and $T^{B_{2u}}_{c0}=8.163$ meV, respectively.
\begin{fullfigure}[t]
  \centering
  \includegraphics[width=0.97\textwidth]{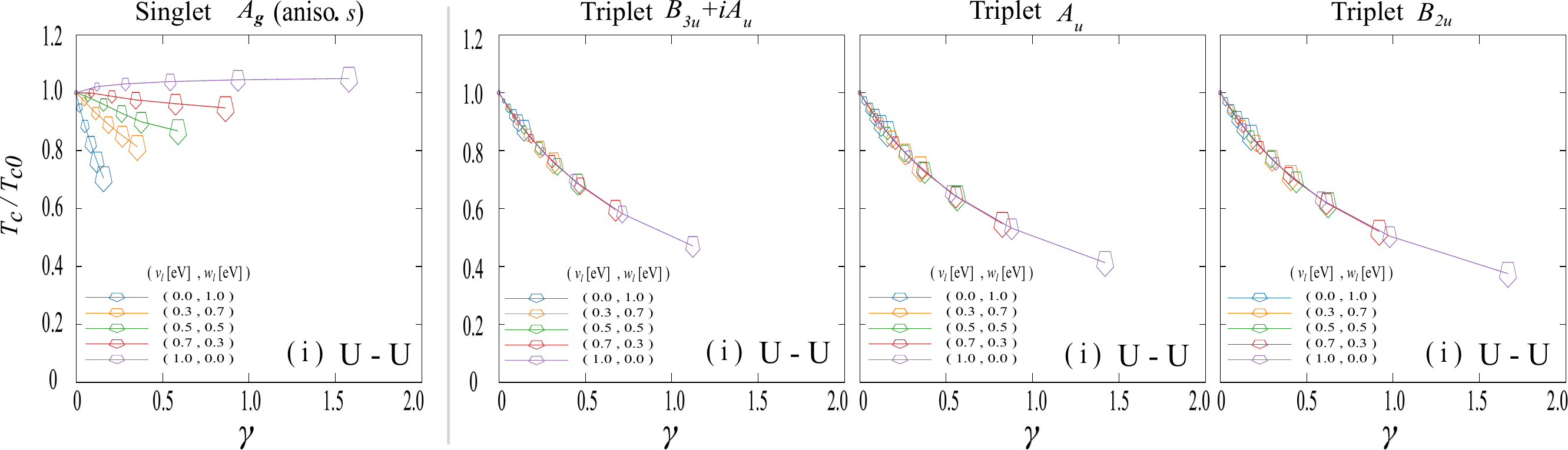}
  \caption{Change in $T_{c}$ as a function of damping $\gamma$ when both nonmagnetic- and magnetic- impurities are introduced. We calculate $(\mathrm{i})$ U-U with $c_{AB}$ = 0.0. The size of the symbols represents to the impurity concentration ($0 \le n \le 0.5\%$).}
  \vspace{-15pt}
  \label{fig5}
\end{fullfigure}
Additionally, we consider three impurity site cases $(\mathrm{i})$ U-U, $(\mathrm{ii})$ U-Te, and $(\mathrm{iii})$ Te-Te and analyze the effects of impurity  correlation and the magnetic properties of impurity site.
The Te impurity site corresponds to the impurity potential of the $p$-orbital.
To simplify the analysis, we ignore the impurity potential of $d$-orbital at the U impurity site and consider only the impurity potential of the $f$-orbital.
Regardless of the type of impurity or impurity site, all the impurity potentials are set to $0 \le v_{l},w_{l}\le 1$ eV, below the bandwidth ($\approx$ 2eV).\cite{Haruna1}

\section{Results and Discussions}
Figure~\ref{fig3} shows the dependences of $T_{c}$ on the impurity concentrations $n$ in the cases of $(\mathrm{i})$ U-U, $(\mathrm{ii})$ U-Te, and $(\mathrm{iii})$ Te-Te.
A situation where only nonmagnetic impurities $(v_{l}=1 \mathrm{eV},w_{l}=0)$ are present in Fig.~\ref{fig3} (A).
In the anisotropic $s$-wave state, $T_{c}$ is kept to be almost constant in accord with the Anderson's theorem, although  $T_{c}$ slightly increases with the increase of impurity concentrations $n$.
The unusual increase in $T_{c}$ will come from the insufficiency of the present analysis scheme, where the damping effect to be actually present is not included in the simplified separable pairing interaction.
Including damping effects in pairing interaction by sophisticated microscopic calculations would suppress the unusual increase of $T_{c}$.
In triplet states, $T_{c}$ decreases with the increase of impurity concentrations $n$ in $(\mathrm{i})$ U-U  and $(\mathrm{ii})$ U-Te,  while $T_{c}$ slightly decreases in $(\mathrm{iii})$ Te-Te, in comparison with the two cases.
The reduction of $T_{c}$ by the contribution of U site reflects  the dominant density of states of U atoms at the Fermi level.\cite{Haruna1}
By contrast, the density of states of Te atom is significantly small at the Fermi level, and therefore impurities at Te atoms do not significantly affect $T_{c}$.
Comparing the four panels in Fig.~\ref{fig3} (A), it is seen that
the off-diagonal components of $c_{AB}$ $(A \neq B)$ between different kinds of impurity sites only negligibly affect $T_{c}$, although they lead to slight increase/decrease of $T_{c}$ for the singlet/triplet states.
Fig.~\ref{fig3} (B) shows a situation where only magnetic impurities $(v_{l}=0,w_{l}=1 \mathrm{eV})$ are present.
For all spin triplet states, the behaviors of $T_{c}$ due to magnetic impurities are qualitatively similar to those due to nonmagnetic impurities.
However, the singlet state exhibits a more pronounced decrease in $T_{c}$.\cite{AG4}
In the singlet state, $T_{c}$ decreases with the increase of magnetic impurity concentrations in (i) U-U and (ii)U-Te, while $T_{c}$ decreases slightly in (iii)Te-Te.
The reduction of $T_{c}$ due to (i) U-U site is larger than that of (ii) U-Te, thus it means that the contribution of U sites strongly influence the reduction of $T_{c}$ in both singlet and triplet states in the case of magnetic impurity.
The gap structures are affected by impurity scattering in three typical cases of $(\mathrm{i})$ U-U are plotted in Fig.~\ref{Fig4}.
In the singlet case with nonmagnetic impurities, the gap magnitude seems to correlate with the $f$-electron weight on the Fermi surfaces (see Fig.~\ref{Fig4}(a)).\cite{fermisurfer}
However, in the singlet case with magnetic impurities, the gap structure tends to be more uniform on the Fermi surfaces. (Fig.~\ref{Fig4}(b)).
In the triplet $A_{u}$ state with nonmagnetic impurities, the impurities modify only the gap magnitude, leaving its momentum dependence unaffected, compared with the pure case (Fig.~\ref{Fig4}(c), the gap in the pure case is not shown here).

We discuss dependence on the relative strength between the nonmagnetic and magnetic impurity potentials.
Here, we express the reduction of $T_{c}$ as a function of $\gamma \equiv T_{c}/T_{n}-1$, where $T_{n}$ is the $T_{c}$ obtained by omitting the second term of Eq.~\eqref{BCS} and therefore $\gamma$ reflects only $T_{c}$ reduction originating purely from the normal self-energy damping.
Here we note variability of the suppression of $T_{c}$ in the triplet states and singlet state.
Whether in singlet or triplet pairing, impurity effects on the normal self-energy lead to quasiparticle damping, which equivalently results in a reduction of $T_{c}$.
However, the way impurities affect the anomalous self-energy differs significantly between singlet $s$-wave and triplet pairings.
In the case of triplet pairings, impurities —whether magnetic or non-magnetic— do not contribute to superconducting correlations at all.
Therefore, the reduction in $T_{c}$ due to the normal self-energy appears directly and follows the AG law irrespective of the magnetic property of impurities, leading to a single curve as a function of $\gamma$ (see Fig.~\ref{fig5}). 
On the other hand, in the $s$-wave superconducting state, non-magnetic impurities assist the superconducting correlations, leading to a recovery of $T_{c}$ (Anderson's theorem).
Magnetic impurities in the $s$-wave superconducting state, however, severely disrupt the superconducting correlations, acting even more destructively than in the triplet case.
As a result, the $s$-wave superconducting state exhibits stronger variability depending on the type of impurity (see Fig.~\ref{fig5}).
So far experimental observations suggest $T_{c}$ suppressions following the AG law.
To ensure consistency with these observations, the following two possibilities are left: (i) triplet pairing or (ii) singlet pairing with magnetic impurities.
Scenario (i) may be naturally acceptable.
However, according to scenario (ii), possibility of the singlet state is not excluded, assumig that magnetic impurities or defects allowing magnetic scattering are unintentionally introduced, possibly due to related magnetic substances, such as U$_{7}$Te$_{12}$.\cite{U7Te12}
Furthermore, according to the results of Th substitution experiment,\cite{Moir2025} the possibility of singlet pairing that is relatively robust against nonmagnetic impurities may still remain.
Thus, elucidating the magnetic properties of impurities is much desired to identify the realized pairing symmetry.

\section{Conclusion}
In conclusion, this study analyzes the effects of impurities in UTe$_{2}$ at zero magnetic field and ambient pressure for various superconducting symmetries.
Although impurity correlation $c_{AB}$ does not make a significant influence, the impurity effect varies significantly depending on the impurity sites.
The main suppression of $T_{c}$ stems from the atomic vacancy at U-sites acting as impurity, while that of Te-sites hardly changes $T_{c}$ except for the case that the anisotropic $s$-wave state of spin-singlet with the nonmagnetic impurity.
Thus, theoretical results for the spin-singlet with magnetic impurities or the triplet states are qualitatively consistent with experimentally observed $T_{c}$ reduction.
To clarify which one is the realistic candidate for the superconducting pairing symmetry of UTe$_{2}$, it is necessary to understand whether the vacancy of U-sites plays a role in nonmagnetic or magnetic impurity.
Identifying the magnetic properties of impurities through experiments such as element substitution is crucial.\cite{Rosa2022,Moir2025}

\section*{acknowledgment}
  We thank Shin-ichiro Shima for technical support with the computer and Shin-ichi Fujimori for valuable discussions.
  This study was supported by the Iketani Science and Technology Foundation and JST SPRING, Japan Grant Number JPMJSP2175.

\nocite{*}
\bibliographystyle{jpsj}
\bibliography{Ref}

\end{document}